\documentstyle[pra,aps]{revtex}
          \draft
          \begin{document}
           \title{A Simple Example of Definitions of Truth,
Validity, Consistency, and Completeness in Quantum Mechanics}
          \author{Paul Benioff\\
           Physics Division, Argonne National Laboratory \\
           Argonne, IL 60439 \\
           e-mail: pbenioff@anl.gov}
           \date{\today}

          \maketitle
          \begin{abstract}  
Besides their use for efficient computation, quantum computers
and quantum robots form a base for studying quantum
systems that create valid physical theories using mathematics 
and physics. If quantum mechanics is universally applicable, then quantum 
mechanics must describe its own validation by these quantum systems. 
An essential part of this process is the development of a coherent theory 
of mathematics and quantum mechanics together. It is expected that such a 
theory will include a coherent combination of  mathematical logical concepts 
with quantum mechanics. 

That this might be possible is shown here by defining truth, validity, 
consistency, and completeness for a quantum mechanical version of a simple
(classical) expression enumeration machine described by Smullyan. Some of the
expressions are chosen as sentences denoting the presence or
absence of other expressions in the enumeration. Two of the
sentences are self referential. It is seen that, for an interpretation based
on a Feynman path sum over expression paths, truth, consistency, and 
completeness for the quantum system have different properties than for the 
classical system. For instance the truth of a sentence $S$ is defined
only on those paths containing $S$.  It is undefined elsewhere.
Also $S$ and its negation can both be true provided they appear
on separate paths. This satisfies the definition of consistency. The
definitions of validity and completeness connect the dynamics of the system to 
the truth of the sentences. It is proved that validity implies consistency. 
It is seen that the requirements of validity and maximal completeness strongly
restrict the allowable dynamics for the quantum system. Aspects of the 
existence of a valid, maximally complete dynamics are discussed. An
exponentially efficient quantum computer is described that is also valid and 
complete for the set of sentences considered here.
         \end{abstract}
         \section{Introduction} 
Most of the activity in quantum computing is supported by the
possibility that some problems can be solved more efficiently on
quantum computers than on classical machines
\cite{Shor,Grover,Lidar1,Lloyd}.  These possibilities in turn have
generated much work towards possible physical realization of
quantum computers using such techniques as NMR \cite{Gershenfeld}
and trapped ions \cite{Cirac}. Other work on theoretical
\cite{Lafl} and experimental \cite{Coryetal} error correction
codes and other methods \cite{Lidar} to make quantum computers
more robust against decoherence resulting from environmental
interference \cite{Zurek} and other influences also is part of
this activity.

The extreme sensitivity of quantum computers to environmental
influences presents a barrier to the practical realization of
quantum computation \cite{Landauer}.  As a result  it is not
clear if quantum computers will ever become a practical reality. 

The same arguments apply to quantum robots \cite{BenQRE}.  These
are mobile quantum systems that include a quantum computer and
other ancillary systems on board that interact with  arbitrary
environments of quantum systems.  The types of environments and
their interactions with quantum robots can be quite general. This
is unlike the case for quantum computers which either seek to
minimize environmental influences or consider very special types
of environments such as oracles \cite{Bennett}, data bases
\cite{Grover}, or additional quantum registers \cite{BenQRE}.  

Another reason for interest in quantum computers and quantum
robots is that they represent a basis for beginning the
description of quantum systems that make decisions, are aware of
their environment, and have important characteristics of
intelligence.  The existence problem for these intelligent
quantum systems is already solved as such systems include the
readers (and hopefully the author) of this paper.

It should be noted that the fact that the only examples of
intelligent quantum systems we know of are macroscopic ($\sim
10^{25}$ degrees of freedom) and may be described classically,
does not remove the need for a quantum mechanical description. By
study of quantum robots or quantum computers one can find out if 
such systems must be essentially classical and, if so, in what
ways a quantum mechanical description fails.

From the viewpoint of this paper an essential activity of
intelligent  systems is the construction of valid physical
theories by use of mathematics and physics.  The details of this
validation process are not important here.  What is
important is that, if the theory being validated is universally applicable, 
then the theory is also  the same theory that describes the dynamics of the 
systems carrying out this validation activity. 

It follows that a universal theory must include a description in some form
of both the mathematical and physical aspects of its own validation. 
This suggests the need for a coherent theory of mathematics and
physics together. Such a theory will  refer to its own
validity and maximal completeness to the maximum extent possible. It also will
be valid and maximally complete.
(The importance of maximal completeness to these ideas was
realized only when the work for this paper was done.) 

If quantum mechanics is universal, then such a coherent theory of
mathematics and quantum mechanics must necessarily include the
description of intelligent quantum systems that can construct and validate 
the coherent theory. As such the coherent theory should  refer to its own 
validity and maximal completeness to the maximum extent possible, and it 
should be valid and maximally complete \cite{PerZur}.

It is to be expected that such a theory will incorporate or
combine aspects of mathematical logic with quantum mechanics.
This would require use in a quantum mechanical context of
mathematical logical concepts such as syntactics and semantics
and their relation to one another
\cite{Shoenfield,Smullyan,Frankel}.  Syntactics deals with
expressions as strings of symbols and languages as sets of
expressions. This includes the description of constants,
variables, terms, formulas, axioms, theorems and proofs.
Semantics is concerned with the meaning of expressions in a
language.  This includes concepts such as interpretations,
models, truth, validity, completeness, and consistency.

There is other work in the literature that recognizes the
potential importance of trying to combine mathematical logical
concepts with quantum mechanics and of describing intelligent
systems in quantum mechanics. The former includes work on
formulas in first order logic \cite{Ozhigov,Buhrman}, set theory
and quantum mechanics \cite{BenZFC,Finkelstein,Svozil}, and other
work \cite{Other}. The latter includes work on consciousness and
quantum mechanics \cite{Penrose,Stapp,Squires}.

In this paper steps will be taken towards the use of mathematical
logical concepts in quantum mechanics by considering a quantum
mechanical system (e.g. head or quantum robot) moving on a lattice of 
stationary quantum systems where
the states of each lattice system are, in general,
linear superpositions of symbol states in some basis.  As the system moves
and interacts with the lattice systems, the system state can be represented
as a linear superposition of symbol string states. If one
symbol is chosen as a blank, then the state corresponds to a
linear superposition of sequences or paths of expressions as sequences of
nonblank symbols separated by one or more blanks. 

The main new feature added here is that some of the expressions 
in the superposition paths will be considered as formal sentences or words 
that are interpreted as having meaning to an outside
observer. This is different than the usual state of affairs where
the outcomes of measurements considered as numbers (symbol
strings) have meaning to the observer carrying out the
experiment.  However, they are not usually considered to be
sentences in some language that may also have meaning to an
outside observer.

It is necessary to be quite clear about this point. This paper
does not address the more ambitious goal of considering a 
quantum system that emits sentences that have meaning to the
quantum system generating the sentences. Here the selection of
which expressions are sentences and how they are to be
interpreted is imposed externally. The quantum system knows
nothing about which expressions are chosen as sentences or how
they will be interpreted.

This is the main reason why the problems first raised by Albert
\cite{Albert,Breuer} are not relevant for this paper. This is the case even
though, as will be seen, the sentences generated will be
interpreted as describing properties of other expressions
generated by the quantum system.

Another point is that, as is well known, computers can be and are
used to manipulate sentences of languages and axiom systems.  An
example is a computer that enumerates the theorems of an axiom
system \cite{Davis}. However all these computer operations deal
with the syntactic properties only of the languages. The fact
that these sentences may or may not have meaning is outside the
realm of what computers, as conceived so far \cite{Penrose}, can
do.

Here the emphasis is on the semantic properties of the language
expressions or their meaning to an external observer.  Following a very 
simple classical example described
by Smullyan \cite{Smullyan} the sentences will be interpreted as
referring to the appearance or nonappearance of other expressions
in the superposition. Based on this interpretation, definitions
of truth, validity, consistency, and completeness for the set of
sentences will be given and some of their properties
investigated.

Since the paper is long, a summary of the sections is in order.  
Following the description of Smullyan's example \cite{Smullyan}  in the
next section, is a description in Section \ref{QMMSM} of a quantum mechanical
model of Smullyan's machine.  The model consists of a quantum
computer or quantum robot moving on a k-ary quantum register as a
1D lattice of k-ary qubytes. (This term is used here instead of
qubits  for values of $k > 2$.) Discrete space and time are
assumed.  The single time step generator for the dynamics of the
overall quantum system  is a unitary operator $T$ acting on the
Hilbert space of system states. A description of the system
components is followed by a description of the properties of $T$.
Various projection operators for expressions and combinations of
expressions are also described. A representation of the overall
state of the evolving quantum enumeration system is given as a
Feynman \cite{Feynman} sum over paths or sequences of
expressions.

In the next and main section \ref{DTV} a simple subset of the set
of sentences in Smullyan's example is chosen. An interpretation
is considered which, for each sentence $S$ in the subset, is
based on the measurement at some time step $n$ for the occurrence
or nonoccurrence of $S$ followed by a later measurement at time
step $n+m$ for the presence or absence of the expression to which
$S$ refers.  

Based on these measurements, definitions are given for the
n,m-truth of the sentences.  The main new feature of the 
interpretation used here is that the n,m-truth of $S$ is
defined only on those paths in the path sum for which $S$
 is present at time $n$. It is undefined on paths not
containing $S$. Among other things, this avoids an
impossible situation that arises in case $S$ appears in
no paths. 

A definition of n,m-true and n,m-false is given for the domain of
definition for the n,m-truth of $S$.  Informally if
$S$ states that some expression $X_{S}$ is present, then $S$ is
n,m-true [n,m-false] if all [not all] paths containing $S$ at
time $n$ contain $X_{S}$ at the later time $n+m$.  If $S$ states
that $X_{S}$ does not appear then $S$ is n,m-true [n,m-false] if
no paths [some paths] containing $S$ at time $n$ contain $X_{S}$
at time $n+m$.

The dependence of these definitions on $n$ and $m$ is problematic
because the n,m-truth of a sentence  is not preserved under for changes in
$m$ or $n$. This problem is removed in the definition of truth
for $S$.  The definition is asymptotic in $n$ in that
it says informally that $S$ is true [false] if it is n,0-true
[n,0-false] in the limit $n\rightarrow \infty$. It turns out that
the limit definition is independent of $m$ so $m$ can be set
equal to $0$.

A definition of n,m-validity is given that is quite similar to
that used in mathematical logic \cite{Shoenfield}.  The
definition connects $T$ to the n,m-truth of $S$ by
saying $T$ is n,m-valid for $S$ if the n-printability of $S$
implies that $S$ is n,m-true. The definition is satisfying in
that it is proved that if $T$ is n,m-valid for $S$ and its
negation then it is n-consistent for $S$ and its negation.  That
is, no path has both $S$ and its negation in the region $[0,n-2]$
of the register lattice.  

The problematic dependence on $n,m$ is removed by a limit
definition of validity. That is, $T$ is valid for $S$ if the
printability of $S$ implies that $S$ is true.  It follows from
this definition that if $T$ is valid for $S$ and its negation
then $T$ is consistent for $S$ and its negation. 

A problem with the definition of validity is that one way for $T$
to be valid for all sentences is to never print any sentences. 
This possibility is removed by the requirement of completeness.
$T$ is complete for $S$ if $S$ is printable (i.e. $S$ appears in
some path). $T$ is complete if $T$ is complete for all sentences;
otherwise it is incomplete. Note that unlike the classical case
both $S$ and its negation can be printable; consistency demands
that they not appear on the same path but says nothing about
their appearance on different paths in the linear superposition.

The set of sentences considered is expanded to include the  self
referential sentence that asserts its own nonprintability and its
negation to show that if $T$ is valid for these two sentences
then they are not printable.  In this case $T$ is incomplete.
This is the equivalent here of G\"{o}del's incompleteness theorem
\cite{Smullyan,Shoenfield}. This suggests the introduction of
maximal completeness; $T$ is maximally complete if it is complete
for all sentences except those excluded by consistency
requirements.

The question of the existence of $T$ that are valid and maximally
complete is discussed in Section \ref{ETVMC}.  The relations between the
truth definitions and correlations between the occcurrence of $S$ and
$X_{S}$ are noted. An exponentially efficient  quantum computer solution 
to the existence problem is shown for a slight generalization of the $T$ 
considered here.

The relation between the existence of a valid, maximally
complete $T$ and the set of expressions taken as sentences is  shown by 
expanding the set of sentences to  include more of those in Smullyan's 
example. It is seen that one must be careful with closed inductive 
definitions which are widely used in mathematical logic.  Some are harmless;
others, such as that used to define sentences in Smullyan's
example, have nontrivial consequences for quantum mechanical systems.

A final discussion section includes the point that there are many
other interpretations possible for the sentences.  An example of
another very simple interpretation is briefly discussed. It is
noted that, due to the freedom in the choice of a basis for
representing the symbols,  many more interpretations are possible
in quantum mechanics than in classical mechanics. 

Another point is that the discussion of the printability both of 
the sentences and the expressions to which they refer may be of
more general applicability than would appear for this special
example. This is based on the fact that any quantum system
telling us something must do so by means of emitting or printing
sentences with meaning. 

\section{Smullyan's Enumeration Machine}
\label{AEM}
Smullyan's example \cite{Smullyan} consists of a (classical)
machine or computer that prints or enumerates expressions
consisting of finite nonempty strings of the five symbols $\sim
\; P \; N \; ( \; )$.  If an expression is printable by the
machine it will eventually be printed. The norm of any expression
$X$ is defined as the expression $X(X)$.

The sentences are defined to be any of the four types of
expressions  $P(X),\; \sim P(X), \;  PN(X), \; \sim PN(X)$ where $X$
is any expression. The sentences are interpeted to apply to the
enumeration generated by the machine in the sense that $P(X)$
means $X$ is printable, $\sim P(X)$ means $X$ is not printable,
$PN(X)$ means the norm of $X$ is printable, and $\sim PN(X)$
means the norm of $X$ is not printable. Thus $P(X)$ is true if
and only if $X$ is printable, $\sim P(X)$ is true if and only if
$X$ is not printable, $PN(X)$ is true if and only if the norm of
$X$ is printable, and $\sim PN(X)$ is true if and only if the
norm of $X$ is not printable. Here and in the following $X$
denotes either an expression variable or a name for a specific
expression.  It should be clear from context which is meant.

Under this interpretation the sentences refer to dynamic
properties of the machine that generates them in that they
describe what the machine does or does not do.  More precisely,
the interpretation is assumed to be {\it valid} for the machine
in that any sentence that is printed is true or, equivalently, no
false sentence is printed. Thus if $P(X)$ is printed, then $X$
has been or will be printed eventually, and if $\sim PN(X)$ is
printed then $X(X)$ will not ever be printed. Similar statements
hold for the other two types of sentences.

The implications, printable implies truth (or falseness implies
not printable), which hold if the interpretation is valid, are
one sided as the converse implications are false. To see this
consider the sentence $\sim PN(\sim PN)$ \cite{Smullyan}.  This
sentence is self referential in that it refers to its own
nonprintability. Thus this sentence is true if and only if it
itself is not printable.  Since the interpretation is supposed to
be valid for the machine, this is a sentence that is true that
the machine cannot print. Also the sentence $PN(\sim PN)$ is not
printable as it is false.

The nonprintability of a true sentence, assuming validity, shows
that for this machine printability is not equivalent to truth of
the sentences.  This is similar to Tarski's Theorem
\cite{Frankel,Smullyan} which says that in any formal axiom
system the set of true formulas is not definable in the system.
Thus  the truth  or falseness of the sentences is a property not
expressible by the machine for the assumed interpretation. In a
similar way the system is incomplete in that neither the sentence
$PN(\sim PN)$ nor its negation are printable.  This is an example of
G\"{o}del's incompleteness theorem \cite{Frankel,Shoenfield} if
printability is interpreted as provability \cite{Smullyan}.

For use in the following two aspects are worthy of note.  One can
show that the sentences $\sim PN(\sim PN)$ and $PN(PN)$ are the
only two self referential sentences. To see this write $\sim
PN(X) = X(X)$ and require that the number of symbols in the
expressions on both sides of the equal sign be the same. This
shows that $X$ must have 3 symbols.  For $PN(X)$ one shows that
$X$ has two symbols.

The other aspect is that, as will be seen later, the definitions
of sentences and their meaning is quite complicated and not
really necessary for the purposes of this paper.  For this reason
all sentences of the form $PN(X),\; \sim PN(X)$ will be excluded
as will sentences of the form $P(X),\; \sim P(X)$ where $X$ is a
sentence. Here sentences will be limited to be of the form
$P(X),\;\sim P(X)$ where $X$ is an expression that is not a
sentence.

\section{A Quantum Mechanical Model of an Enumeration Machine} 
\label{QMMSM} 
\subsection{Component Description}
A quantum mechanical model of a  symbol enumeration machine as
described above consists of a multistate head or quantum robot
moving on a lattice or quantum register of 5-ary qubytes. The
interaction between the quantum robot and the lattice qubytes is
local and includes changing the states of the neighborhood
qubytes.  For the purposes of this paper it is immaterial whether
the whole system is regarded as a multiregister quantum computer
or as a quantum robot or as a multistate head moving on a quantum
register \cite{BenQRE}. 

The set of 5 symbols represented by the states of each qubyte
are $\sim , \; P, \; (, \; ),$ and $0$. The $0$ denotes the blank symbol
and will be interpreted as a spacer to separate a string of 
5 symbols into a string of expressions separated by spacers. A convenient 
set of basis states for the quantum register is the set of states
$|\underline{s}\rangle = \otimes_{j=-\infty}^{\infty}
|s_{j}\rangle$ where $|s_{j}\rangle$ denotes site $j$ qubyte in a state
corresponding to  any one of the 5 symbols.
The state $|\underline{s}\rangle$ describes an infinite symbol string state 
for which at most a finite number of symbols are nonblank. This limitation,
referred to as the $0$ state tail condition, is used to keep the Hilbert 
space of the overall system, including the register, separable.

The states of the head or quantum robot can be represented in the
form $|l,j\rangle$ where $|l\rangle$ denotes any of the $L$
states of the internal degrees of freedom of the head and
$|j\rangle$ is the lattice position state of the head. For
example if the internal degrees of freedom of the head or quantum
robot consist of another  $m$ state head moving on a lattice of
$n$ qubits $L=mn(2^{n})$. Based on the above a general normalized
state of the overall system has the form $\Psi
=\sum_{l,j,\underline{s}} c_{l,j,\underline{s}}
|l,j,\underline{s}\rangle$ where the $c_{l,j,\underline{s}}$ are
arbitrary complex coefficients whose absolute squares sum to
unity. The $\underline{s}$ sum is over all lattice system basis
states that satisfy the $0$ state tail condition.

\subsection{System Dynamics}
\label{SD}
The dynamics of the overall system is given by a unitary step
operator $T$ that represents the changes occurring in a single
time step. If $\Psi (0)$ is the overall system state at time $0$
then $\Psi (n)=T^{n}\Psi (0)$ is the state at time $n$.

In order that $T$ describe  enumerations of symbols on the qubyte
lattice it is necessary to require that the states of qubytes in
local lattice regions become asymptotically  (as $n\rightarrow
\infty$) stationary.  Any dynamics in which the states of local
regions of the quantum register are always changing does not
represent an enumeration. Mathematically this condition can be
expressed by the requirement that the expectation value $\langle
\Psi (n)|P_{\underline{s}_{R}}|\Psi (n)\rangle$ has a limit as
$n\rightarrow \infty$.  Here $P_{\underline{s}_{R}}$ is the
projection operator for the symbol string state
$|\underline{s}\rangle$ in a local region R of the lattice.

To keep things simple this requirement will be satisfied here by
requiring $T$ to describe motion of the quantum robot in one
direction only on the 1D lattice of qubytes. Each iteration of
$T$ will move the quantum robot or head one site to the right.
During this motion the internal state of the head and the states
of the qubytes at the original and final locations of the head
can be changed.

To this end let $T$ be given by
\begin{equation}
T= U\otimes u. \label{Tdef}
\end{equation}
Here $u$ is the unitary shift operator for moving the quantum
robot one lattice site to the right and $U$ is an arbitrary  
$25L$ dimensional unitary operator on the $L$ internal head states
and states of the two lattice qubytes, one at and one just to the
right of the head location.

The action of $T$ on each overall system state $|l,j,\underline
{s}\rangle$ is given by 
\begin{equation}
T|l,j,\underline{s}\rangle = |j+1,\underline{s}_{\neq [
j,j+1]}\rangle \sum_{l^{\prime},s^{\prime}_{j},s^{\prime}_{j+1}} 
|l^{\prime},s^{\prime}_{j},s^{\prime}_{j+1}\rangle \langle
l^{\prime},s^{\prime}_{j},s^{\prime}_{j+1}|U|l,s_{j},s_{j+1}\rangle
\end{equation}
where $ |\underline{s}_{\neq [j,j+1]}\rangle$ is the state of lattice qubytes  
outside of sites $j,j+1$ and $u|j\rangle = |j+1\rangle $ has been used.

To be consistent with the $0$ state tail condition and the choice of $0$ as
the symbol blank or vacuum, the initial state of interest here is
$|0,0,\underline{0}\rangle$. This state has the head in internal state $0$
at lattice site $0$ and a completely blank quantum register.  Inclusion of
initial wave packet states of different head positions and internal states
is not necessary here.

An expansion of $T^{m}$ acting on the initial state
$|0,0,\underline{0}\rangle$, in terms of
intermediate states gives
\begin{eqnarray}
T^{m}|00\underline{0}\rangle & = & \vert m, \underline{0}_{\neq
[0,m]}\rangle \sum_{s_{0},\cdots ,s_{m}}\sum_{s^{\prime}_{1},\cdots ,
s^{\prime}_{m}}\sum_{l_{1},\cdots ,
l_{m}} \vert l_{m},s_{[0,m-1]},s^{\prime}_{m}\rangle \langle l_{m},
s_{[0,m-1]},s^{\prime}_{m}|U| l_{m-1}, s_{[0,m-2]}s^{\prime}_{m-
1},0_{m}\rangle \nonumber \\ \mbox{} & \times &  \langle l_{m-1},
s_{[0,m-2]},s^{\prime}_{m-1},0_{m}|U| l_{m-2}, s_{[0,m-
3]}s^{\prime}_{m-2},0_{[m-1,m]}\rangle \cdots  \langle l_{1},s_{0},s^{\prime}_{1},
0_{[2,m]}|U| 0,0,\underline{0} \rangle.
\end{eqnarray}
Here $s_{[0,b]}$ denotes a string of symbols in the region
$0,1,\cdots , b$ of lattice sites. 

This equation shows that the quantum
system with dynamics given by $T$ is a satisfactory enumeration
system in that once the head passes a lattice site the states of
all qubytes in passed regions, denoted by the states $|s_{[0,b]}\rangle$ are
not changed by more iterations of $T$. The growth in the passed lattice region by
one site per $T$ iteration is shown by the increase of $b$ to $b+1$ in the
states $|s_{[0,b]}\rangle$ appearing in each matrix element. The sum over
the unprimed $s$ represents the completed effect of the passage of the head 
in generating the states $|s_{[0,b]}\rangle$; the sum over primed $s$ gives 
intermediate changes in qubyte states at the location of the head.

\subsection{Some Expression Projectors}
\label{BAEP}
Expressions $X$ are defined as consecutive finite strings of any
of the four symbols $P,\; (,\; ),\; \sim$ with $0$ excluded
within $X$.  To separate $X$ from other expressions in a symbol
string, it is required that the terminal symbol of $X$ is
followed by at least one $0$.  Similarly the initial symbol of
$X$ is preceded by at least one $0$.  Let $a,b$ be the initial
and terminal symbol site location of $X$ where $l(X)$ is the
length of $X$.  One has $a=b-l(X)+1$.  

Define $Q_{X,b}$ to be the projection operator for finding $X$ at
lattice sites $a,a+1,\cdots ,b$, $0s$ at sites $a-1,b+1$, and any
symbol, including the blank, at other lattice sites.  The quantum
robot can be anywhere and in any internal state. This operator is
basic to all that follows.

Let $[m,n]$ with $n>m$ be a lattice region of sites $m,m+1,\cdots
,n$.  Define the projection operator $Q_{X,[m,n]}$ to be the
projection operator for finding $X$ anywhere in the region
$[n,m]$. $Q_{X,[m,n]}$ is defined by
\begin{equation}
Q_{X,[m,n]} = \sup_{k=m+l(X)-1}^{n}Q_{X,k}
\label{Qdef}\end{equation}
 $Q_{X,[m,n]}$ is the 
projector for  $X$ being somewhere in the region $[m,n]$. The
least upper bound is used because  $Q_{X,k}$ and $Q_{X,k^{\prime}}$ with
$k\neq k^{\prime}$ are not necessarily orthogonal. 
$Q_{X,[m,n]}=0$ if the region is too short to accomodate $X$.

Let $Q_{\neg X,[m,n]}$ be the projection operator for not finding
$X$ anywhere in $[m,n]$.  Clearly
\begin{equation}
Q_{X,[m,n]} + Q_{\neg X,[m,n]} =1.
\end{equation}
Additional useful properties of these projectors are
\begin{eqnarray}
Q_{X,[m,n]} & < & Q_{X,[m,n+1]} \nonumber \\ 
Q_{\neg X,[m,n]} & > & Q_{\neg X,[m,n+1]} \label{Qltgt}
\end{eqnarray}
and 
\begin{equation}
Q_{X,[m,n]} =\sum_{j=l(X)}^{n}Q^{1st}_{X,[m,m+j]}. \label{Q1st}
\end{equation}
Here $Q^{1st}_{X,[m,m+j]}$ is the projection operator for $X$
occurring in the region $[m,m+j]$ with the terminal symbol of $X$
at site $m+j$,  $0$s at sites $m+j+1,\; m+j-l(X)$, and no $X$ anywhere else in
$[m,m+j]$. $Q^{1st}_{X,[m,m+j]}$ is the identity on all other
lattice sites.  The superscript $1st$ denotes the fact that $X$ does not
occur in intervals $[m,m+k]$ with $k<j$, i.e. the first occurrence of $X$.
 Note that the projection operators in the $j$ sum are pairwise orthogonal.

In what follows projection operators are needed for expressions,
that, once generated by iterations of $T$, are not changed by
further iterations. One way to achieve this is to include
projection operators for head positions to the right of
expressions of interest. To this end let $Q^{h}_{X,[m,n]}$ be the
projection operator for $X$ anywhere in the region $[m,n]$ where
$X$ is separated from other expressions in the region by one or
more $0s$. If $X$ ends at site $n$ there is a $0$ at site $n+1$;
if $X$ begins at site $m$, there is a $0$ at site $m-1$. The head
is at site $n+2$. That is 
\begin{equation}
Q^{h}_{X,[m,n]} =Q_{X,[m,n]}Q^{h}_{n+2} \label{QhXn}
\end{equation}
where $Q^{h}_{n+2}$ is the projection operator for the head at
position $n+2$ and in any of the $L$ internal states. A useful
generalization is the projection operator $Q^{h}_{X,[m,n],k}
=Q_{X,[m,n]}Q^{h}_{n+k+2}$ for $X$ anywhere in the interval
$[m,n]$ and the head k+2 sites beyond $n$. For nonnegative $k$
this projector has the following commutation relation with $T$ as
defined by Eq. \ref{Tdef}:
\begin{equation}
TQ^{h}_{X,[m,n],k} = Q^{h}_{X,[m,n],k+1}T. \label{Qcomm}
\end{equation}

The limit projector $Q^{h}_{X}$ defined by 
\begin{equation}
Q^{h}_{X}= \sum_{n=0}^{\infty} Q^{h}_{X,[0,n]} \label{QhX}
\end{equation}
corresponds to $X$ located anywhere to the right of the origin
and the head at least $2$ sites beyond the terminal symbol of
$X$. The projectors in the sum are pairwise orthogonal because of
the orthogonality of the head location projectors.  Whether or
not this limit exists is not important here because the limit
operator can always be replaced by $Q^{h}_{X,[0,n]}$ for some $n$
in any of the matrix elements that occur in this work. 

Let $X$ and $Y$ be two expressions. Then
\begin{equation}
Q^{h}_{X\wedge Y,[m,n]}=Q^{h}_{X,[m,n]}Q^{h}_{Y,[m,n]}
\label{Qand} \end{equation}
is the projection operator for finding $X$ and $Y$ in the region
$[m,n]$ and the head at site $n+2$.  This projection operator is
zero if the region is too small to contain $X$ and $Y$ without
overlap.  This is the case even if $Y$ is a subexpression of $X$
because the projectors include blank symbols preceding and
following each expression.

More generally let $Q^{h}_{\wedge_{k=1}^{m}Y_{k}}$ be defined
by
\begin{equation}
Q^{h}_{\wedge_{k=1}^{m}Y_{k}}=\Pi_{k=1}^{m}Q^{h}_{Y_{k}}
=\sum_{j=0}^{\infty}Q_{\wedge_{k=1}^{m}Y_{k},[0,j-
2]}Q^{h}_{j}. \label{Qhprod}
\end{equation}
This is the projection operator for finding expressions
$Y_{1},\cdots Y_{k}$ anywhere to the right of the lattice origin
and to the left of the head (and a $0$ just left of the head).

It is important to note that the symbols $\neg ,\; \wedge$ appearing in the 
subscripts are in the metalanguage used to describe the properties of the 
system. They do not appear in the expressions or sentences described here.
The operator $Q^{h}_{\neg X}$ projects out all expression path states that
do not contain $X$ anywhere. This is the case whether $X$ is or is not a
sentence.

\subsection{Sums over Paths of Expressions}
\label{SOPE}

At this point it is worthwhile to look more closely at iterations
of $T$ and the generation of an exponentially growing tree of
paths or sequences of expressions separated by strings of $0s$.
To this end define the projection operators   
\begin{eqnarray}
Q_{0} & = & \sum_{j=-\infty}^{\infty}Q_{0,j-1}Q_{j}^{h} \nonumber
\\
Q_{\neq 0} & = & \sum_{j=-\infty}^{\infty}Q_{\neq 0,j-1}Q_{j}^{h}
\end{eqnarray}
Here $Q_{j}^{h}$ is the projector for the head at site $j$,
$Q_{0,j-1}$ is the projector for a $0$ at site $j-1$, and $Q_{\neq
0,j-1}$ is the projector for any one of the 4 symbols
$(,),P,\sim$ at site $j-1$. It is clear that $Q_{0} + Q_{\neq 0}
= 1$.

These operators can be used to separate $T$ defined by Eq.
\ref{Tdef} into the sum of two operators $T_{0},\;T_{\neq 0}$:
\begin{equation}
T = (Q_{0}+Q_{\neq 0})T = T_{0} + T_{\neq 0}.
\end{equation}
The projectors are chosen so that for any overall system state
$\Psi$, $T_{0}\Psi$ and $T_{\neq 0}\Psi$ show respectively a $0$
or an expression symbol state for the qubyte at the site last
visited by the head. These states have the property that
iteration of $T$ on these states does not change these qubyte
states as changes are limited to those qubytes at and immediately
to the right of the head location. Iteration of $T_{0}$ generates a spacer 
state as a finite string of $0s$, and iteration of $T_{\neq 0}$ generates 
a linear superposition of expression states.  Note that $T_{0}$ 
and $T_{\neq 0}$ do not commute.

Using the fact that $T^{n} = (T_{0}+T_{\neq 0})^{n}$ and collecting together
powers of $T_{0}$ and $T_{\neq 0}$ gives
\begin{eqnarray}
T^{n} & = & \sum_{t=1}^{n} \sum_{h_{1},\cdots ,h_{t}}^{\delta (\sum ,n)} [(T^{h_{t}}_{\neq
0}T^{h_{t-1}}_{0}T^{h_{t-2}}_{\neq 0} \cdots T^{h_{2}}_{0}T^{h_{1}}_{\neq 0} +
T^{h_{t}}_{0}T^{h_{t-1}}_{\neq 0} \cdots T^{h_{2}}_{\neq
0}T^{h_{1}}_{0})\delta_{t,\mbox{odd}} \nonumber \\
& & \mbox{} + (T^{h_{t}}_{0}T^{h_{t-1}}_{\neq 0}\cdots T^{h_{1}}_{\neq 0} + 
T^{h_{t}}_{\neq 0}T^{h_{t-1}}_{0}\cdots T^{h_{1}}_{0})\delta_{t,\mbox{even}}]. \label{Talt}
\end{eqnarray}
The $t$ sum is over the number of expressions and intervening spacers, and the $h$ sums are over the length of the
expressions  and spacers in an alternation with $t$ spacers and expressions.  
The upper limit of the $h$ sums denotes the requirement that $h_{1}+h_{2}+ 
\cdots +h_{t} =n$. The above shows that four types of alternations are
possible; they may begin or end with either $T_{0}$ or $T_{\neq 0}$. How they 
end depends on how they begin and whether $t$ is even or odd.  The two types 
for $t$ odd are shown in the first line above multipled by a delta 
function for $t$ odd. The second line gives the two types for $t$ even.

The above can be used to expand $T^{n}|0,0,\underline{0}\rangle$ into a
Feynman sum \cite{Feynman} over sequences or paths of expression states
separated by $0s$ similar to the sum over phase paths used elsewhere
\cite{BenQRE}.  In order to keep things simple the expansion will be given
for the first alternation type only with $t$ odd and full account will be 
taken of the fact that once the head passes a lattice region no further 
interactions occur with the qubytes in the region.  One has
\begin{eqnarray}
T^{n}|0,0,\underline{0}\rangle & = &
\sum_{t=1}^{n}\sum_{h_{1},\cdots ,h_{t}=1}^{\delta (\sum
,n)}\sum_{l_{1},\cdots ,l_{t}}\sum_{s^{\prime}_{1},\cdots , s^{\prime}_{t}} 
\sum_{X_{1},X_{2},\cdots , X_{(\frac{t+1}{2})}}
|l_{t},n,\underline{0}* X_{1}*\underline{0}*
X_{2}*\underline{0}*,\cdots ,* \underline{0}*
X_{(\frac{t+1}{2})}*s^{\prime}_{t}*\underline{0}\rangle \nonumber \\
& & \mbox{} \times \langle l_{t},X_{[(t+1)/2]}*s^{\prime}_{t}| T^{h_{t}}_{\neq
0}|l_{t-1},s^{\prime}_{t-1}*\underline{0}\rangle  \langle l_{t-
1},\underline{0}*s^{\prime}_{t-1}| T^{h_{t-1}}_{0}|
l_{t-2},s^{\prime}_{t-2}*\underline{0}\rangle \nonumber \\
& & \mbox{}\cdots  \langle
l_{2},\underline{0}*s^{\prime}_{2}| T^{h_{2}}_{0}|
l_{1},s^{\prime}_{1}*\underline{0}\rangle\langle
l_{1},X_{1}*s^{\prime}_{1}|T^{h_{1}}_{\neq
0}|0,0,\underline{0}\rangle. \label{pathsum}
\end{eqnarray}

The $l$ sums are over head or robot internal states, and the $s^{\prime}$
sums are over intermediate states (including $0$) of qubytes at the head 
location. As shown earlier they may be changed by the next iteration of $T$. 
The $X$ sums are over all possible completed expressions of length specified 
by the $h$ sum terms. The head position state has been suppressed in the 
matrix elements. The asterisk denotes concatenation of symbols and expressions.

Each matrix element shows the state changes resulting from one alternation.
For example $T^{h_{j}}_{0}$ is active on the lattice region extending from
$\sum_{k=1}^{j-1}h_{k}$ (the initial head position) to $\sum_{k=1}^{j}h_{k}$
(the final head position). It converts the state of qubytes in the lattice
region from $|s^{\prime}_{j-1}*\underline{0}\rangle$ to $|\underline{0}
*s^{\prime}_{j}\rangle$ where $\underline{0}$ denotes a string of $h_{j}$
$0s$ and a sum over $s^{\prime}_{j}$ is implied. (Here the subscript $j$ is
an alternation index, not a lattice site.)

The action of $T^{h_{j}}_{\neq 0}$ differs only in that the final state is
$|X_{j}*s^{\prime}_{j}\rangle$ where $X_{j}$ is an expression of length
$h_{j}$.  Sums over $s^{\prime}_{j}$ and $X_{j}$ are implied. Note that this
represents the generation of a linear superposition of length $h_{j}$
expression states in the specified lattice lattice region.  Except for
future entanglements with states of qubytes not yet reached by the head, the
expression states are not
changed by more iterations of $T$.  This is shown by the fact that each
$X_{j}$ also appears in the final output state $|l_{t},n,\underline{0}* 
X_{1}*\underline{0}*X_{2}*\underline{0}*,\cdots ,* \underline{0}*
X_{[\frac{t+1}{2}]}*s^{\prime}_{t}*\underline{0}\rangle$ which shows the
head in the internal state  $l_{t}$ at position $n$.  The
state  shows the sequence of expressions $X_{1},X_{2},\cdots
,X_{[\frac{t+1}{2}]}$ separated by finite strings of $0s$. The
terminal  expression $X_{[\frac{t+1}{2}]}$ is complete for those
components in which one more iteration of $T$ converts $s^{\prime}_{t}$
to $0$.  It is incomplete in the other components of the sums of
Eq. \ref{pathsum}. The qubytes at lattice sites not in the region
$[0,n]$ are in state $|0\rangle$.

Eq. \ref{pathsum} shows a sum over all expression sequence states 
consistent with the requirement that the lengths of all
expressions and spacers sum to $n$.  It includes components for
one expression of length $n-2$ up to components for $[n/2]$
expressions each containing one symbol. Other components with
expressions of varying length are also included.

The amplitude associated with each expression sequence state
$|l_{t},n,\underline{0}* X_{1}*\underline{0}*
X_{2}*\underline{0}*,\cdots ,* \underline{0}*
X_{[\frac{t+1}{2}]}*s^{\prime}_{t}*\underline{0}\rangle$ is given by a
partial sum over products of matrix elements  shown in Eq.
\ref{pathsum}. It depends sensitively on the properties of the
$25L$ dimensional unitary operator $U$, Eq. \ref{Tdef}, which
shows the changes in the lattice system states as the quantum
robot moves along the lattice.

It is clear from the above that each expression path may contain
many different expressions $,X,Y,Z,\cdots$ as well as repetitions
of expressions. Let $Q^{h}_{Y,[0,n-2]}$ be the projection
operator for the expression $Y$ appearing somewhere in the
region $[0,n-2]$ and the head at site $n$
in any internal state, Eq. \ref{QhXn}. Then the state
$Q^{h}_{Y} T^{n}|0,0,\underline{0} \rangle =
Q^{h}_{Y,[0,n-2]}T^{n}|0,0,\underline{0}\rangle$, expanded as a
sum over expression paths (Eq. \ref{pathsum}), contains all
expression path states with $Y$ appearing somewhere in the region
$[0,n-2]$. For later times $m+n$ the state
$T^{m}Q^{h}_{Y,[0,n-2]}T^{n} |0,0,\underline{0}\rangle$, expanded
as a path sum, shows a sum over expression path states with $n+m$
symbols, including blanks,  that have in common the appearance of
$Y$ somewhere in the region $[0,n-2]$.

\section{Truth, Validity, Consistency, and Completeness}
\label{DTV}

\subsection{Sentences and Their Interpretation}
\label{STI}
So far expressions have been discussed without any mention of which ones are
sentences and how the sentences should be interpreted.  Here sentences are
those expressions with the form $P(X),\; \sim P(X)$ where $X$ is any 
expression that is not a sentence. In the following $S$ will often be used 
to denote a sentence and $X_{S}$ the expression to which the sentence refers.
 
A sentence $S$ is defined to be n-printable if $\langle 0,0,\underline{0}
|(T^{\dagger})^{n}Q^{h}_{S}T^{n}|0,0,\underline{0}\rangle > 0$. It is not n-printable if
the matrix element equals $0$.  $S$ is printable if if
$\lim_{n\rightarrow \infty}\langle 0,0,\underline{0}|(T^{\dagger})^{n}Q^{h}_{S}
T^{n}|0,0,\underline{0}\rangle > 0$. Otherwise it is not printable. 
$Q^{h}_{S}$ is given by Eq. \ref{QhX} with $X=S$.  The limit clearly exists
because the matrix element is nondecreasing as $n$ increases and
is real, positive and bounded above by $1$.

The definition of printability means that if $S$ is printable it
will appear somewhere in the sum of Eq. \ref{pathsum}. The limit
$n\rightarrow \infty$ is needed because, in general, there is no upper bound
on when a sentence must first appear in the different paths. 
Conversely a sentence is not printable if it never appears in any
path. 

There are many possible ways to interpret the sentences $S$. Here the
interpretation of $P(X)$, stated informally,  is that all paths in the 
path sum of Eq. \ref{pathsum} that contain $P(X)$ also contain $X$. The
informal interpretation of $\sim P(X)$ is that no path containing $\sim
P(X)$ also contains $X$. $P(X)$ or $\sim P(X)$ are true if their
informal meaning statements are true.  The goal of the following
is to make precise these informal expressions of the meaning and
truth of the sentences.

A new feature introduced by these interpretations is that the
meaning of a sentence $S$ is limited to those paths containing
$S$.  $S$ has no meaning in paths not containing $S$. It
says nothing about the presence or absence of $X_{S}$ in these
paths. This is a consequence of the presence of many paths in the
path sum of Eq. \ref{pathsum}.  Classically, where there is just
one path, this partial definability is reflected in the fact that, for the
interpretation used here, any sentence not appearing in the enumeration 
path also has no meaning for the expressions in the path.  

The meaning of $S$ is closely connected to the carrying out of
measurements on the quantum enumeration system. Suppose one
carries out a measurement of the observable $SQ^{h}_{S} + \neg
SQ^{h}_{\neg S}$ at time $n$ on the state $\Psi(n) =
T^{n}|0,0,\underline{0}\rangle$ and $m$ time steps later carries
out a measurement of the observable $X_{S}Q^{h}_{X_{S}}+ \neg
X_{S}Q^{h}_{\neg X_{S}}$ on the component
state $Q^{h}_{S}|\Psi(n)\rangle$. This can be described by adding
two ternary qubytes to the quantum enumeration systems to record
the outcomes of these operations. The qubyte states
$|i\rangle_{S} ,\; |i\rangle_{X_{S}}$ denote the initial or no
measurement state and the other states denote the possible two
outcomes of each measurement. 

The overall process is described by unitary operators $U^{S}$ and
$U^{X_{S}}$ that establish correlations between the states of the
enumeration system and the qubytes.  The result of carrying out
the two measurements is given by 
\begin{eqnarray}
U^{X_{S}}T^{m}U^{S}\Psi(n) |i\rangle_{S} |i\rangle_{X_{S}} & = &
Q^{h}_{X_{S}}T^{m}Q^{h}_{S}\Psi(n) |1\rangle_{S}
|1\rangle_{X_{S}} \nonumber \\
& & \mbox{} + Q^{h}_{\neg X_{S}}T^{m}Q^{h}_{S}\Psi(n)
|1\rangle_{S} |0\rangle_{X_{S}} + T^{m}Q^{h}_{\neg S}\Psi(n)
|0\rangle_{S} |i\rangle_{X_{S}}. \label{SXSmeas}
\end{eqnarray}

In the above the amplification or decoherence by interaction with
the environment \cite{Zurek} necessary to complete the
measurement process is ignored as it is not needed here.  In
Peres' language \cite{Peres} the above is a premeasurement.

The limitation of the $X_{S}$ measurement to component states
containing $S$ is made because extension of the measurement of
$X_{S}$ to component states not containing $S$ is not needed
here. However, if desired, the extension can be included. The only effect is
to add an additional term to Eq. \ref{SXSmeas}.  No conclusions are affected.

In order to keep things simple it has been assumed in the above
that the projection operators $Q^{h}_{X}$ and $Q^{h}_{\neg X}$
where $X$ is an expression or sentence can be measured in one
time step.  This is unrealistic as the measurement involves
searching for $X$ in an arbitrarily large region of the lattice. 
A more realistic approach would be to replace this one step
operation with a measurement described as a multistep search task
carried out by a quantum robot moving along the lattice of
qubytes \cite{BenQRE}.  For each time $n$ the search would
terminate as a finite lattice region $[0,n-2]$ only needs to be
searched.  However the number of steps in the search is
polynomial in $n$.

Eq. \ref{SXSmeas} clearly supports the idea that for a
measurement at time $n$ the meaning or interpretation of $S$ is
limited to those component path states in which $S$ occurs
somewhere in the region $[0,n-2]$. $S$ has no meaning for those
component states in which $S$ occurs nowhere in $[0,n-2]$.  This
would remain the case even if the measurement of $X_{S}$ were
extended to paths not containing $S$. The equation also shows
that for the measurement at time $n$ the domain of states for
which $S$ has meaning is the Hilbert space spanned by the set of
normalized states $T^{m}Q^{h}_{S}\Psi(n)/\|
Q^{h}_{S}|\Psi(n)\rangle \|$ for $m=0,1,\cdots$. This Hilbert
space is empty if $S$ has more than $n-2$ symbols.

This Hilbert space is also the state  space or domain of truth
definition for the sentence $S$ at time $n$. It corresponds
to the states shown in the first two righthand terms of Eq.
\ref{SXSmeas}.  In other words, the truth at time $n$ for a sentence $S$
is definable on the (unnormalized) states $T^{m}Q^{h}_{S}\Psi
(n)$ for $m=0,1,\cdots$. It is not definable on the states
$T^{m}Q^{h}_{\neg S}\Psi (n)$.  

Many definitions of n,m-truth for the sentences are possible. 
Here n,m-truth will be defined as follows:  $S=P(X)$  is
 n,m-true [n,m-false] if the amplitude of the second
righthand term of Eq. \ref{SXSmeas} $=0\; [>0]$.  $S=\sim
P(X)$ is n,m-true [n,m-false] if the amplitude of the first
righthand term $=0 \; [>0]$.  That is
\begin{eqnarray}
P(X) \begin{array}{l} \mbox{ is n,m-true } \\ \mbox{ is n,m-false }
\end{array} & \mbox{ if } & \langle \Psi
(n)|Q^{h}_{P(X)}(T^{\dagger})^{m}Q^{h}_{\neg
X}T^{m}Q^{h}_{P(X)}|\Psi (n)\rangle \begin{array}{l} = 0 \\ > 0
\end{array} \label{PXntr} \\
\sim P(X) \begin{array}{l} \mbox{ is n,m-true } \\ \mbox{ is
n,m-false } \end{array} & \mbox{ if } & \langle \Psi (n)|Q^{h}_{\sim
P(X)}(T^{\dagger})^{m}Q^{h}_{ X}T^{m}Q^{h}_{\sim P(X)}|\Psi
(n)\rangle   \begin{array}{l} = 0 \\ > 0 \end{array}
\label{notPXntr}
\end{eqnarray}

These definitions follow the practice in mathematical logic
\cite{Shoenfield} of defining truth of formal sentences relative to the
informal truth of the statements that are the interpretation of the
sentences. Here the interpretation of $P(X)$ is the mathematical statement
$\langle \Psi (n)|Q^{h}_{P(X)}(T^{\dagger})^{m}Q^{h}_{\neg
X}T^{m}Q^{h}_{P(X)}|\Psi (n)\rangle =0$ that describes a property of the
quantum system shown by the measurement of Eq. \ref{SXSmeas}. $P(X)$ is
n,m-true [n,m-false] if the mathematical statement is true [false] for the
system.  Similarly $\sim P(X)$ is n,m-true [n,m-false] if the statement
$\langle \Psi (n)|Q^{h}_{\sim P(X)}
(T^{\dagger})^{m}Q^{h}_{X}T^{m}Q^{h}_{\sim P(X)}|\Psi (n)\rangle =0$ is
true [false] for the system.

The above definitions also have different properties from the usual 
definitions.  Besides the need for a state domain of truth
definition, it is {\it not} the case that $P(X)$ is n-true if
$\sim P(X)$ is false and conversely.  This is a consequence of
the quantum mechanical nature of the enumeration system with the
presence of multiple paths in the path sum.  In fact there is no
reason why these sentences cannot both be n,m-true on their
respective domains of definition.

An equivalent definition of n,m-truth follows from the fact that
$Q^{h}_{X}$ and $Q^{h}_{\neg X}$ are orthogonal and sum to unity:
\begin{eqnarray}
P(X) \begin{array}{l} \mbox{ is n,m-true } \\ \mbox{ is n,m-false }
\end{array} & \mbox{ if } & \langle \Psi (n) |
(T^{\dagger})^{m}Q^{h}_{X} T^{m}Q^{h}_{P(X)}|\Psi (n)\rangle   -
\langle \Psi (n)|Q^{h}_{P(X)}|\Psi (n)\rangle \begin{array}{l} =
0 \\ < 0 \end{array} \label{PXn2tr} \\
\sim P(X) \begin{array}{l} \mbox{ is n,m-true } \\ \mbox{ is
n,m-false } \end{array} & \mbox{ if } & \langle \Psi (n)|(T^{\dagger})^{m} 
Q^{h}_{\neg X}T^{m}Q^{h}_{\sim P(X)}|\Psi
(n)\rangle   - \langle \Psi (n)|Q^{h}_{\sim P(X)}|\Psi (n)\rangle
\begin{array}{l} = 0 \\ < 0 \end{array} \label{notPXn2tr}
\end{eqnarray}
The fact that the projectors $Q^{h}_{S}$ and
$(T^{\dagger})^{m}Q^{h}_{X_{S}}T^{m}$ commute has been used to obtain these
expressions.

The connection of these definitions of m,n-truth to the
measurement is shown by the fact that the definition correlates
the n,m-truth of $S$ to the amplitudes of states appearing in the
first two terms of Eq. \ref{SXSmeas}.  Also the domains of
definability and undefinability of n,m-truth are shown in the equation.

These definitions also illustrate another reason for use of a
domain of definability for the truth of a sentence at time $n$.
For the special case where the probability of finding $S$ at time
$n$ is $0$ then the normalized measurement state corresponding to
finding $S$ at $n$, $Q^{h}_{S}\Psi (n)/\|Q^{h}_{S}\Psi (n)\|$ is
undefined. In addition, if the definitions of 
n,m-truth also applied for this case, it would be impossible for $S$ to be
n,m-false.  This untenable situation is avoided by the use of
domains of definability.

A problem with these definitions is their dependence on $m$ and
$n$. To see this let $T$ be such that there is no upper bound on
the distance of first appearance of $X_{S}$ from the origin. 
That is, for each $m$ there exist paths in Eq. \ref{pathsum} such
that for some $m^{\prime} >m$ $X_{S}$ is not in $[0,n+m-2]$ but is in
$[0,n+m^{\prime} -2]$. This means that $P(X)$ can be $n,m$-false
and $n,m^{\prime}$-true.   Also $\sim P(X)$ can be n,m-true
and $n,m^{\prime}$-false.  This follows from Eq. \ref{Qltgt}.

It is also the case that n,m-truth depends on $n$.  For example  $T$ may be 
such that $S$ is $n,m$-true but is $n^{\prime},m$-false for some
$n^{\prime} >n$. This is possible because as $n$ increases, extensions and
branching of paths occurs. Since $S$ may first occur in some of these
extensions or branches, it is possible for $S$ to  be $n^{\prime},m$-false 
on some of these extensions. Additional dependence on $n$ arises from the
above argument for $m$ dependence because the size of the region for the
$X_{S}$ measurement, Eq. \ref{SXSmeas}, depends on both $n$ and $m$.
 
The definition of truth should be such as to exclude the
dependence on $n$ and $m$. It should also maximize the domain of
definability of truth for a sentence. These goals are attained by
taking the limit  $n\rightarrow \infty$ in Eq. \ref{SXSmeas} and
in the definitions of n,m-truth.  The desired maximization is
achieved because  $\langle \Psi (n)|Q^{h}_{S}|\Psi (n)\rangle$ is
a nondecreasing function of $n$ bounded above by $1$. 

In addition, for the limit definition, the value of $m$ is
arbitrary as the limit values of the matrix elements are
independent of $m$. This is shown in the Appendix. Because of
this the value of $m$ will be chosen to be $0$. Based on these
aspects true and false are defined as follows for the two types
of sentences:
\begin{eqnarray}
P(X) \begin{array}{l} \mbox{ is true } \\ \mbox{ is false }
\end{array} & \mbox{ if } & \lim_{n\rightarrow \infty} \langle \Psi 
(n)| Q^{h}_{\neg X}Q^{h}_{P(X)}|\Psi (n)\rangle \begin{array}{l}
= 0 \\ > 0 \end{array} \label{PXtr} \\
\sim P(X) \begin{array}{l} \mbox{ is true } \\ \mbox{ is false }
\end{array} & \mbox{ if } & \lim_{n\rightarrow \infty} \langle \Psi
(n)| Q^{h}_{ X}Q^{h}_{\sim P(X)}|\Psi (n)\rangle  
\begin{array}{l} = 0 \\ > 0. \end{array} \label{notPXtr}
\end{eqnarray}

Equivalent definitions corresponding to those of Eqs.
\ref{PXn2tr} and \ref{notPXn2tr} are:
\begin{eqnarray}
P(X) \begin{array}{l} \mbox{ is true } \\ \mbox{ is false }
\end{array} & \mbox{ if } & \lim_{n\rightarrow \infty} \langle \Psi
(n)| Q^{h}_{X}Q^{h}_{P(X)}|\Psi (n)\rangle   -\lim_{n\rightarrow
\infty} \langle \Psi (n)|Q^{h}_{P(X)}|\Psi (n)\rangle
\begin{array}{l} = 0 \\ < 0 \end{array} \label{PX2tr} \\
\sim P(X) \begin{array}{l} \mbox{ is true } \\ \mbox{ is false }
\end{array} & \mbox{ if } & \lim_{n\rightarrow \infty} \langle \Psi
(n)|Q^{h}_{\neg X}Q^{h}_{\sim P(X)}|\Psi (n)\rangle  
-\lim_{n\rightarrow \infty} \langle \Psi (n)|Q^{h}_{\sim
P(X)}|\Psi (n)\rangle \begin{array}{l} = 0 \\ < 0 \end{array}
\label{notPX2tr}
\end{eqnarray}
These definitions remove the $n$ dependence in that if a sentence
$S$ is n,0-true for each $n$ then it is true. The converse,
namely that if $S$ is true, it is n,0-true for each $n$, is true
for $S=\sim P(X)$.   This follows from Eq. \ref{notPXtr} as the
matrix element is nondecreasing as $n$ increases. The converse is
not necessarily true for $S=P(X)$.  However one can show from Eq.
\ref{Q1st} and the material in the Appendix that if $P(X)$ is
true then it is n,$\infty$-true for each $n$.

\subsection{Validity and Consistency}
\label{VC}
The definitions of n,m-truth and truth for the sentences given
above clearly depend on the system dynamics $T$.  However they
place no restrictions on $T$.  Each sentence with a nonempty
domain of n,m-truth definition (i.e. n-printable) can be n,m-true
or n,m-false, (or true or false) on the domain. The main idea
here is to use validity to connect $T$ to the truth of the
sentences.

Informally the idea is that $T$ is n,m-valid for a sentence $S$
if $S$ is in fact n,m-true on its domain of definition. That is,
all paths containing $P(X)$ [$\sim P(X)$] in the region $[0,n-2]$
also contain [do not contain] $X$. This is captured in the
following definition: \\
\\ {\bf definition 1:} {\em $T$ is {\em n,m-valid} for a sentence $S$ if $S$ is 
not n-printable or $S$ is n-printable and n,m-true on its domain of
definition.} \\ 
\\
An equivalent statement of the definition is that $T$ is
n,m-valid for $S$ if n-printability of $S$ implies that $S$ is
n,m-true on its domain of definition. As noted the domain is
nonempty if and only if $S$ is n-printable. 

This definition is quite similar to that used in mathematical
logic where a sentence is valid for some interpretation if it is
is true for the interpretation \cite{Shoenfield,Frankel}. (It is
also similar to the notion of accuracy or correctness used by
Smullyan \cite{Smullyan}.) The main difference is that here the
domain of definability of n,m-truth plays an important role. Note
that for $T$ as defined here no sentence with more than $n-2$
symbols is n-printable. However, $T$ is n,m-valid for all these sentences.

The definition of n,m-validity for $S$ is closely connected with
the measurement of $S$ and $X_{S}$, Eq. \ref{SXSmeas}. If $S$ is
n-printable and $S=P(X)$ $[S=\sim P(X)]$, then the n,m-validity
of $T$ for $S$ means that the second [first] righthand term of
Eq. \ref{SXSmeas} is $0$.

The equation also shows that $T$ is n,m-valid for $S$ if the
probability that $S$ does not appear in the region $[0,n-2]$ plus
the probability that $S$ appears in the region and is n,m-true on
its domain of definition equals $1$. In terms of matrix elements
for the two types of sentences this is expressed by:
\begin{eqnarray}
\mbox{For $P(X)$ }\:  \langle \Psi (n)| (T^{\dagger})^{m} Q^{h}_{X}
T^{m}Q^{h}_{P(X)}|\Psi (n)\rangle & + & \langle \Psi(n) |Q^{h}_{\neg P(X)}|\Psi
(n)\rangle  = 1 \nonumber \\ \mbox{For $\sim P(X)$ }\: \langle \Psi (n)|
(T^{\dagger})^{m} Q^{h}_{\neg X}T^{m}Q^{h}_{\sim P(X)}|\Psi (n)\rangle & + &
\langle \Psi(n) |Q^{h}_{\neg \sim P(X)}|\Psi (n)\rangle
 =1 \label{TnvalS}
\end{eqnarray}

The definition of n,m-validity for $S$ can be extended to all
sentences.  $T$  is {\it n,m-valid} if it is n,m-valid for all
sentences $S$.  In terms of single sentence measurements
described by Eq. \ref{SXSmeas} if $T$ is n,m-valid then Eqs.
\ref{TnvalS} hold for each sentence.  Measurements can be
limited to just those sentences that will fit in the region
$[0,n-2]$ as $T$ is automatically n,m-valid for all other
sentences because they are not n-printable.
 
It may be possible to combine all the single sentence
measurements into one measurement observable and all $X_{S}$
measurements into another observable. If this is the case then
n,m-validity can be described in terms of one type of
measurement.  These observables are much more complex than those
in Eq. \ref{SXSmeas}.  The complexity is shown even for the
measurement for just two sentences $S$ and $S^{\prime}$. In this
case the left hand side of Eq. \ref{SXSmeas} is replaced by
$U^{X_{S^{\prime}}}U^{X_{S}}
T^{m}U^{S^{\prime}}U^{S}\Psi(n)|i\rangle_{S^{\prime}}
|i\rangle_{S} |i\rangle_{X_{S}}|i\rangle_{X_{S^{\prime}}}$.  The
right hand side is a sum of nine terms, four for components
containing both $S$ and $S^{\prime}$ as each sentence can be n,m,-true or
n,m-false, two for each of the two components
containing just one of the two sentences in which n,m-truth is defined for
the sentence appearing, and one for components
containing neither sentence. For this term, n,m-truth is undefined
for both sentences.  Note that the two unitary operators
for $X_{S}$ and $X_{S^{\prime}}$ commute as do the two for $S$
and $S^{\prime}$.

If $T$ is n,m-valid for both these sentences, then the n,m-truth conditions
give the result that at most four
of the the nine terms are nonzero.  These are the terms that exclude
paths showing the n,m-falseness of either sentence.  
As a specific example let $S=P(X)$ and $S^{\prime}=\sim P(Y)$. If
$T$ is n,m-valid for these two sentences, then only four terms 
may be nonzero. The term of most intererst here is that containing all paths
in which both $P(X)$ and $\sim P(Y)$ appear and are n,m-true.  This is the
state $Q^{h}_{X\wedge \neg Y}T^{m}Q^{h}_{P(X)\wedge \sim P(Y)}\Psi(n)
|1\rangle_{P(X)} |1\rangle_{\sim P(Y)}|1\rangle_{X}|0\rangle_{Y}$. Conservation
of probability and the unitarity of $T$ gives the result:

\begin{displaymath}
\langle \Psi (n)|(T^{\dagger})^{m}Q^{h}_{X\wedge \neg
Y}T^{m}Q^{h}_{P(X)\wedge \sim P(Y)}|\Psi (n)\rangle  = 
\langle \Psi (n)|Q^{h}_{P(X)\wedge \sim P(Y)}|\Psi (n)\rangle.
\end{displaymath}

This equation is interesting because if $X=Y$
then the lefthand matrix element is $0$ because the projection
operator $Q^{h}_{X\wedge \neg X} =0$.  Thus the following
theorem has been proved: \\
\\ {\bf theorem 1:} {\em Let $T$ be n,m-valid for $P(X)$ and for $\sim P(X)$. 
Then $\langle \Psi(n)|Q^{h}_{P(X) \wedge \sim P(X)}|\Psi(n)\rangle
=0$.} \\ 
\\
This theorem is satisfying since it shows that if $T$ is
n,m-valid for the sentences $P(X)$ and $\sim P(X)$, then these
sentences are n-consistent for $T$.  Here a sentence and its
negation are defined to be {\it n-consistent} for $T$ if no path
in the path sum of Eq. \ref{pathsum} contains both the sentence
and its negation in the lattice region $[0,n-2]$.
That is $\langle \Psi(n)|Q^{h}_{P(X)\wedge \sim P(X)}|\Psi
(n)\rangle =0$.  

It is clear that n,m-validity of $T$ for $P(X)$ and $\sim P(X)$
is a sufficient condition for $T$ to be n-consistent for these
sentences because without the condition of n,m-validity for $T$
there is no reason why a sentence and its negation could not both
appear in some path. It is not a necessary condition since it is
possible for $T$ to be n-consistent but not n,m-valid for these two sentences. 
Note that n,m-validity of $T$ for $P(X)$ and $\sim P(X)$ does not prevent them 
from appearing in some paths (i.e. both can be n-printable). They cannot, 
however, both appear on the same path.

The dependence of n,m-truth on $n$ and $m$ also results in a similar
dependence for n,m-validity. For example if
$S=P(X)$ then $T$ may be $n,m^{\prime}$-valid for $S$ but not n,m-valid for
$S$ where $m^{\prime} >m$. If $S=\sim P(X)$ then $T$ may be n,m-valid for
$S$ but not $n,m^{\prime}$-valid for $S$.  A similar
dependence holds for changes in $n$.

As was done for n,m-truth, this undesirable n,m-dependence can 
be removed by setting $m=0$ and taking the limit $n \rightarrow \infty$.  
To this end one defines $T$ to be valid for $S$ by: \\
\\ {\bf definition 2:} {\em $T$ is {\em valid at $S$} if either $S$ is not 
printable or $S$ is printable and true on its domain of truth definition.} \\
\\
An equivalent statement of the definition is that $T$ is valid
for $S$ if printability imples truth on the domain of truth
definition. Here printability is defined as in subsection
\ref{STI} by $\lim_{n\rightarrow \infty} \langle \Psi
(n)|Q^{h}_{S}|\Psi (n)\rangle >0$ and truth by Eqs. \ref{PXtr}
and \ref{notPXtr} (or Eqs. \ref{PX2tr} and \ref{notPX2tr}). 

The definition of validity for a sentence can be extended to all
sentences by defining $T$ to be valid if $T$ is valid for all
sentences.  That is $T$ is valid if for all sentences $S$
printability of $S$ implies that $S$ is true on its domain of
definition. This definition can be given in an equivalent form
based on Eq. \ref{TnvalS}: \\
\\ {\bf definition 3:} {\em $T$ is {\em valid} if for all expressions $X$ that 
are not sentences, {\em 
\begin{eqnarray}
 \lim_{n\rightarrow \infty}\langle \Psi
(n)|Q^{h}_{X}Q^{h}_{P(X)}|\Psi (n)\rangle & + &
\lim_{n\rightarrow \infty} \langle \Psi(n) |Q^{h}_{\neg
P(X)}|\Psi (n)\rangle  =1 \nonumber \\ 
\lim_{n\rightarrow \infty}\langle \Psi (n)| Q^{h}_{\neg
X}Q^{h}_{\sim P(X)}|\Psi (n)\rangle & + & \lim_{n\rightarrow \infty}
\langle \Psi(n)|Q^{h}_{\neg \sim P(X)}|\Psi (n)\rangle  =1 \label{Tval}
\end{eqnarray} }} \\
\\
Note that the limit $n\rightarrow \infty$ commutes with the sum
in Eq. \ref{Tval}.

The definition of n-consistency can also be extended to the limit
$n\rightarrow \infty$ as follows: \\
\\ {\bf definition 4:} {\em $T$ is {\em consistent for $P(X)$ and $\sim P(X)$} 
if {\em $\lim_{n\rightarrow \infty} \langle \Psi (n)|Q^{h}_{P(X)\wedge
\sim P(X)}|\Psi (n)\rangle =0$.}} \\ 
\\
Since the matrix element in this definition is nondecreasing as
$n$ increases, this definition is equivalent to defining $T$ to
be consistent for $P(X)$ and $\sim P(X)$ if for each $n$, $T$ is
n-consistent for these two sentences.

It should be noted that the use of consistency here seems
different from that used in the consistent histories theory of
quantum mechanics \cite{Griffiths,Gell,Omnes}. Here consistency
refers to properties of certain pairs of sentences generated by a
quantum system. In the consistent histories approach consistency
is a property of projectors on a tensor product of Hilbert spaces 
associated with multitime events or histories of a quantum system
\cite{Griffiths}.

One can use the above definition of consistency  and the
definition of validity to obtain the following theorem: \\
\\ {\bf theorem 2:} {\em If $T$ is valid for $P(X)$ and $\sim P(X)$ then $T$ 
is consistent for $P(X)$ and $\sim P(X)$.} \\ \\
To prove this theorem note that by the properties of the
projection operators the following two relations hold:
\begin{eqnarray*}
\lim_{n\rightarrow \infty}\langle \Psi
(n)|Q^{h}_{X}Q^{h}_{P(X)}Q^{h}_{\sim P(X)}|\Psi (n)\rangle & \leq
& \lim_{n\rightarrow \infty}\langle \Psi (n)|Q^{h}_{X}
Q^{h}_{\sim P(X)}|\Psi (n)\rangle \\ \lim_{n\rightarrow \infty}
\langle \Psi (n)|Q^{h}_{\neg X}Q^{h}_{P(X)}Q^{h}_{\sim P(X)}|\Psi
(n)\rangle & \leq &\lim_{n\rightarrow \infty} \langle \Psi
(n)|Q^{h}_{\neg X} Q^{h}_{P(X)}|\Psi (n)\rangle .
\end{eqnarray*}
Since $T$ is valid for these two sentences, by Eqs. \ref{PXtr}
and \ref{notPXtr}, the righthand sides of these two inequalities
are both equal to $0$. Addition of these two inequalities and
noting that the limits commute with the sum, and using $Q^{h}_{X}
+Q^{h}_{\neg X} =1$ gives $\lim_{n\rightarrow \infty}\langle \Psi
(n)|Q^{h}_{P(X)}Q^{h}_{\sim P(X)}|\Psi (n)\rangle =0$ which
proves the theorem.

$T$ is said to be consistent if for all expressions $X$ that are
not sentences, $T$ is consistent for $P(X)$ and $\sim P(X)$. It
is clear from the above that if $T$ is valid then it is
consistent.

\subsection{Completeness}
As noted if $T$ is valid then each sentence is either not
printable or it is printable and true.  Thus any $T$ which does
not print any sentence at all is valid.  Such $T$ are easy to
construct as it is easily decidable which expressions are
sentences.

Completeness is used to remove this possiblity: \\
\\ {\bf definition 5:} {\em $T$ is {\em complete for a sentence $S$} if $S$ is 
printable. $T$ is {\em complete} if it is complete for all sentences. 
Otherwise it is incomplete.} \\
\\
As is known from the G\"{o}del incompleteness theorem,
\cite{Shoenfield,Smullyan,Frankel} there exist sets of sentences
with axioms that are incomplete.  The same situation can occur
here.  To see this add to the set of sentences (of the form
$P(X)$ and $\sim P(X)$) the two sentences, in Smullyan's example,
$PN(X)$ and $\sim PN(X)$ where $X=\sim PN$.  For this value of
$X$ $\sim PN(X)$ is self referential. 

For  the interpretation of the sentences used here (and addition of $N$ to
the set of expression symbols) it is easy to
prove the following theorem: \\
\\ {\bf theorem 3:} {\em Let $T$ be valid for the sentences $\sim PN(\sim PN)$ 
and $PN(\sim PN)$.  Then neither of these sentences is printable.} \\ \\
To prove this theorem assume $\sim PN(\sim PN)$ is printable. 
Since $T$ is valid for this sentence it must then be true.  By
the definition of truth, Eq. \ref{notPXtr}, $ \lim_{n\rightarrow
\infty} \langle \Psi (n)|Q^{h}_{\neg \sim PN(\sim PN)}Q^{h}_{\sim
PN(\sim PN)}|\Psi (n)\rangle =\lim_{n\rightarrow \infty}\langle
\Psi(n)|Q^{h}_{\sim PN(\sim PN)}|\Psi (n)\rangle$. Since the two
projection operators appearing in the left hand matrix element
are orthogonal, the matrix element is zero for each $n$.  Thus
both matrix elements in the equality must equal $0$ which
contradicts the assumption that $\sim PN(\sim PN)$ is printable. 
So this sentence is not printable. 

For the sentence $PN(\sim PN)$ the properties of the projection
operators  give the result $\lim_{n\rightarrow \infty} \langle
\Psi (n)| Q^{h}_{PN(\sim PN)}Q^{h}_{\sim PN(\sim PN)}|\Psi
(n)\rangle \leq  \lim_{n\rightarrow \infty}\langle \Psi
(n)|Q^{h}_{PN(\sim PN)}|\Psi (n)\rangle$. Since $T$ is valid for
this sentence, then either the right hand matrix element $=0$ or
it is $>0$ and equals the left hand matrix element (by the
definition of truth. But, by  theorem 2 on consistency,
this is impossible. Thus the right hand matrix element is zero in
either case,  so $PN(\sim PN)$ is also not printable, which
proves the theorem.

It follows that for this expanded set of sentences $T$ is not
complete.  Note that this holds for all $U$ in the definition of 
$T$, Eq. \ref{Tdef}. This suggests that one define a concept of
maximal completeness.  A valid $T$ is {\it maximally complete} if
it is complete for all sentences subject only to the requirements
of consistency.  For the interpretation considered here for the
sentences, all uses of consistency including theorem proofs,
ultimately depend on the fact that the projection operators
$Q^{h}_{X}$ and $Q^{h}_{\neg X}$ are orthogonal for any
expression $X$.

The other self referential sentence, $PN(PN)$ (Section \ref{AEM}), can be 
included at no cost.  The reason is that if it is printable it is trivially 
true.  Thus the requirement that $T$ be valid for $PN(PN)$ imposes no
restrictions on $T$.  Completeness for $PN(PN)$ just requires
that the sentence be printable.

\section{The Existence of Valid and Maximally Complete $T$}
\label{ETVMC}

The above considerations raise the issue of the existence of step
operators (or generators of the dynamics) $T$ that are valid and
maximally complete for the set of sentences and interpretation
considered here.  This is not trivial because it is clear from
the above that these requirements are quite restrictive. (For the
set of sentences used here maximal completeness is equivalent to
completeness.)

It is an open question if there exist $T$ satisfying Eq. \ref{Tdef} that are
valid and complete for the sentences considered here. One aspect is that the 
requirement that the sentences be true requires the presence of strong 
correlations between the occurrences of a sentence $S$ and the expression 
$X_{S}$. If $S=\sim P(X)$ then $\sim P(X)$ is true if and only if there 
is a complete negative correlation between the occurrence of $\sim P(X)$ and 
$X$ no matter how far apart they are.  $P(X)$ is true if and only if there is 
a complete positive correlation between the occurrence of $P(X)$ and $X$.  
However, unlike the case for $\sim P(X)$, this correlation can be of finite 
length because once $X$ occurs in a path containing $P(X)$ the truth definition 
is satisfied for all paths that are extensions of the path segment 
containing $X$. 

These correlation requirements are shown by Eqs. \ref{PX2tr} and
\ref{notPX2tr}. In essence, these equations show the deviations from 
the condition of no correlation imposed by the truth definitions. No
correlation is expressed by
\begin{displaymath}
\lim_{n\rightarrow \infty} \langle\Psi (n)|Q^{h}_{X_{S}}Q^{h}_{S}|\Psi
(n)\rangle = \lim_{n\rightarrow \infty}\langle \Psi (n)|Q^{h}_{X_{S}}|\Psi
(n)\rangle \langle \Psi (n)|Q^{h}_{S}|\Psi (n)\rangle
\end{displaymath}

If one relaxes the requirement that $T$ satisfy Eq. \ref{Tdef} to allow
limited backward motion, then there exist both classical and quantum
computer solutions to the existence question. The quantum computer (or
robot) proceeds as follows:  For each $n=1,2,\cdots ,2^{M}$, generate the
superposition $1/\sqrt{2}(|0\rangle_{a}+|\sim\rangle_{a})
|P(X)0\rangle_{[a+1,b]}$ of states of all $P(X),\sim P(X)$ where $X$ is any 
length $n$ expression.  Then for each $X$ correlate whether $X$ is [is not] a 
sentence (a decidable property) with the states $|1\rangle_{q} [|0\rangle_{q}]$ 
of a qubit $q$. Change an adjacent
state of n $0s$ to $X$ only for the sentence $P(X)$ and only for those $X$ that 
are not sentences. Increase $n$ by $1$ and move to the next region of 
$2(n+1)+5$ blank sites.  

For each $n$ the overall state transformations are given by
\begin{eqnarray}
|0\rangle_{q} & |\underline{0}\rangle_{[a,b]} & |\underline{0}\rangle_{[b,c]} 
 \longrightarrow  \frac{1}{\sqrt{4^{n}}}\sum_{X=1}^{4^{n}} |0\rangle_{q}
\frac{(|0\rangle_{a}+|\sim\rangle_{a})}{\sqrt{2}} 
|P(X)0\rangle_{[a+1,b]}|\underline{0}\rangle_{[b,c]} \nonumber \\
& &\mbox{} \longrightarrow \frac{1}{\sqrt{4^{n}}} (\sum_{X}^{\neg S}|0\rangle_{q} +
\sum_{X}^{S}|1\rangle_{q})\frac{(|0\rangle_{a}+|\sim \rangle_{a})}{\sqrt{2}} 
|P(X)0\rangle_{[a+1,b]}|\underline{0}\rangle_{[b,c]} \nonumber \\
& \longrightarrow & \frac{1}{\sqrt{4^{n}}}\sum_{X}^{\neg S}|0\rangle_{q}
\frac{(|0P(X)0\rangle_{[a,b]}|X\rangle_{[b,c]} +|\sim
P(X)0\rangle_{[a,b]}|\underline{0}\rangle_{[b,c]})}{\sqrt{2}}
+\sum_{X}^{S}|1\rangle_{q}\frac{(|0\rangle_{a}+|\sim\rangle_{a})}{\sqrt{2}} 
|P(X)0\rangle_{[a+1,b]}|\underline{0}\rangle_{[b,c]} \label{qucom}
\end{eqnarray}
Here $b=n+4-a$ and $c=n-1-b$.  The sums $\sum_{X}^{\neg S}, \sum_{X}^{S}$ are  
over all length $n$ expressions that are not ($\neg S$) or are ($S$) sentences.
The last line shows that nothing is done to the $\sum_{X}^{S}$ components.  
The number of length $n$ expressions that are not sentences is $4^{n}-\Delta$  
where $\Delta =4^{n-3}+4^{n-4}$. The 
quantum robot starts at position $a$ in internal state $|0\rangle$ and ends
in state $|0\rangle$ at position $c+1$ to repeat the cycle for $n+1$.  The
value of $M$ reflects the presence of a quantum computer with a register of
at least $M$ qubits on board the quantum robot. In this way the 
quantum robot has with it a record of the active value of $n$. 

This quantum computer is exponentially efficient in that 
the number of steps required to generate the final state for a $T$ that is
valid and complete for all sentences $S$ where the length of $X_{S}$ is
$\leq 2^{M}$ is polynomial in $(2^{M}+1)2^{M}/2$. The efficiency is shown by
the fact that the number of sentences
included in the final state is given by $N=\sum_{n=1}^{2^{M}}2(4^{n}-\Delta)$.

This efficiency is lost if one wants to determine by  measurement of $X$
values if $T$ is in fact valid and complete as $\approx N$ repetitions of the 
preparation and measurement of the state shown above would have to be carried 
out.  A much more promising approach is to carry out a Fourier transform over 
the qubytes in region $[a+3,b-1]$ that give the argument $X$ of $P(X),\sim P(X)$ 
in the $\sum_{X}^{\neg S}$ final state part of Eq. \ref{qucom}. This gives the state 
\begin{displaymath}
\frac{1}{4^{n}\sqrt{2}}\sum_{Y=0}^{4^{n}-1}\sum_{X}^{\neg S}|0\rangle_{q} 
e^{2\pi iYX/{4^{n}}}(|0P(Y)0\rangle_{[a,b]}|X\rangle_{[b,c]} +|\sim
P(Y)0\rangle_{[a,b]}|\underline{0}\rangle_{[b,c]}).
\end{displaymath}
Here the probability distribution of $0P(Y)$ as a function of
$Y$ is completely different from that of $\sim P(Y)$. For $0P(Y)$
the distribution is uniformly distributed over all $Y$ whereas for $\sim
P(Y)$ the distribution is to a good approximation (of order
$(\Delta /4^{n})^{2}$) a delta function at $Y=0$. (Here, depending on context,
expressions are either strings of n symbols or 4-ary numbers). Since these
probability distributions are so different it is likely that, as is the case
for other quantum algorithms, \cite{Shor,Lidar1}, they can be
determined to good accuracy in polynomially (in $n$) many repetitions of
preparation and measurement of $0P(Y)$ and $\sim P(Y)$.
 
This quantum computer solution for the existence problem refers to an example
for which the same quantum system generates both
the sentences and the expressions to which the sentences refer. Of more
general interest is the case where the quantum system generating the sentences 
is distinct from the quantum system to which the sentences refer. This is the 
usual case in physics where the system carrying out measurements is distinct 
from the system being measured. Study of these systems is deferred to future
work.

It is worth noting that the existence problem is strongly related to the set
of expressions admitted as sentences. Suppose, following Smullyan 
\cite{Smullyan}, one expands the set of
sentences by dropping the requirement that $X$ is not a sentence. 
This introduces many complexities into the discussion.  Suppose
for example $T$ is valid and maximally complete for all sentences
in the expanded set.  Then the sentence $P(P(X))$ is printable
and true which means that $P(X)$ is printable and also true. This
means informally  that all paths containing $P(P(X))$ also
contain $P(X)$ and all paths containing $P(X)$ contain $X$.  Of
course $P(X)$ may be present in paths not containing $P(P(X))$.

Application of the same argument to $P(\sim P(X))$ gives the
result that all paths containing this sentence must contain $\sim
P(X)$ and none of the paths containing $\sim P(X)$ may contain
$X$.  In a similar fashion, none of the paths containing $\sim
P(P(X))$ may contain $P(X)$ and all paths containing $P(X)$
contain $X$; for $\sim P(\sim P(X))$ no path containing this
sentence may contain $\sim P(X)$ and no path containing $\sim
P(X)$ may contain $X$.

This is a complex set of requirements especially because each of
the $8$ sentences involved is printable.  For example consistency
means that $P(P(X))$ and $\sim P(P(X))$ have no paths in common. 
The same holds for the pair $P(\sim P(X))$ and $\sim P(\sim
P(X))$.  However in addition consistency demands that $P(P(X))$
and $P(\sim P(X))$ also have no paths in common. The reason is
that any path containing both these sentences must contain both
$P(X)$ and $\sim P(X)$ which is not possible.  The same argument
fails for the pair $\sim P(P(X))$ and $\sim P(\sim P(X))$ because
validity and completeness mean that any path containing both
these sentences must contain neither $P(X)$ nor $\sim P(X)$. 
This is possible. 

This shows how the complexity of the requirements of validity and
completeness for $T$ grows if one includes sentences of order
greater than the first order, the atomic sentences, which is the
set considered here. As is shown above the complexity is
appreciable even for the eight types of second order sentences
described above.

This also shows quite forcefully that closed inductive
definitions, which are used so much in mathematical logic, should
be used here only with careful examination of the consequences. 
To see the problem, note that Smullyan's definition of sentences 
\cite{Smullyan} restricted to sentences of the type $P(X)$ and $\sim P(X)$ 
can be given as:
\begin{enumerate}
\item All expressions $P(X)$ and $\sim P(X)$ where $X$ is not a
sentence are sentences (the atomic sentences). 
\item If $X$ is a sentence so are $P(X)$ and $\sim P(X)$. 
\item Sentences are only as defined above.
\end{enumerate}

The problem with this definition resides in the second
requirement which expresses closure.  Here a definition in terms
of inductive orders is more suitable.  For each $k+1$ {\it
candidate} sentences of order $k+1$ are defined as those
expressions of the form $P(X)$ or $\sim P(X)$ where $X$ is a
sentence of order $k$. For each  $k =1,2,\cdots$ $T$ is k-valid
and k-complete if it is valid and complete for all sentences of
order $\leq k$. If and only if there exist $T$ that are k-valid
and k-complete should one consider expanding the set to include
the candidate sentences of order $k+1$ as sentences.

The reason for this is that as the sentence order increases the
validity and completeness requirements become increasingly
onerous.  For example there may well be many $T$ that are k-valid
and k-complete but are not (k+1)-valid and (k+1)-complete.  An
example of a $T$ that is valid and complete for the first order
(atomic) sentences but is not valid and complete for the second
order sentences would be any $T$ that generates paths containing
both $P(P(X))$ and $P(\sim P(X))$ and is valid and complete for
the atomic sentences.

This situation is even more complex if sentences of the form
$PN(X)$ and $\sim PN(X)$ are admitted (Section \ref{AEM}) even if
$X$ is not a sentence.  Not only must one deal with the fact that
different sentences denote the same expression (e.g $P(Y)$ and
$PN(X)$ where $Y=X(X)$), but for some $X$ these sentences
generate chains of sentences. A very simple chain of length 2
results from $X =P$.  If $PN(P)$ is printable and true so is
$P(P)$. Since $P(P)$ is a sentence it is also printable and true.
The chain stops here as $P$ is not a sentence.

Other $X$ give longer chains. For $X=PN(P$ one has 
\begin{displaymath}
PN(PN(P)\rightarrow PN(P(PN(P)\rightarrow  P(P(PN(P(
P(PN(P)\rightarrow P(PN(P( P(PN(P
\end{displaymath}
The chain terminates because the last expression is not a
sentence [no terminal $)$]. The chain for $X=PN\sim P$ 
\begin{displaymath}
PN(PN(\sim P)\rightarrow PN(\sim P(PN(\sim P)\rightarrow \sim P(
(PN(\sim P(\sim P(PN(\sim P)
\end{displaymath}
stops one stage earlier than the chain for $X=PN(P$ because the
last sentence, which is true by validity asserts the
nonprintability of an expression.

There is even a nonterminating chain. To see this set $X=PN(PN$.  
The first few terms are
\begin{displaymath}
PN( PN(PN)\rightarrow  PN( PN(PN(PN)\rightarrow
PN(PN(PN(PN(PN(PN) \rightarrow \cdots
\end{displaymath}
It is clear that the number of symbols in the successive
sentences grows exponentially with increasing position in the
chain.  If $N_{n}$ denotes the number of repetitions of $PN$ in
the $nth$ position then one sees that $N_{n+1} = 2(N_{n}-1)$
where $N_{1} =3$.

\section{Discussion}
\label{Disc}

It is important to reemphasize that the choice of which
expressions are sentences and the particular interpretation
assumed for the sentences is external to the quantum enumeration
system.  It is imposed from the outside. As such the restrictions
that validity and completeness place on the dynamics are relative
to this interpretation. The quantum system is completely silent
on which expressions, if any, are sentences and how they are to
be interpreted.  This is the case even for $T$ that are valid and
maximally complete.  It is a very long way from valid and
complete $T$ as described here to the dynamics of quantum systems
that describe to the maximum extent possible their own validity
and completeness.

Nevertheless the example described here has aspects that may be
useful for a description of systems that describe their own
validity and completeness.  It is expected, for instance that the
definitions of validity, completeness, and possibly consistency will remain.  
So will some aspects of the definition of truth. It also may be that the
notion of printability will remain in more general systems.  This
follows from the fact that any quantum system that is telling us
something about some other quantum system has to print or
enumerate strings of symbols as some type of sequence of physical
signals.  It also has to tell us which groups of signals (i.e.
expressions are meaningful (sentences) and which are meaningless
noise. If the system cannot print or enumerate anything it cannot
tell us anything.

It should also be noted that there are many other interpretations
of the sentences in addition to the interpretation described
here.  A very simple one defines printability of expressions as
is done here but defines the truth of sentences differently. 
That is $P(X)$ is true [false] if $\lim_{n\rightarrow \infty}
\langle \Psi(n)| Q^{h}_{X}|\Psi(n)\rangle >0\; [=0]$ and $\sim
P(X)$ is true [false] if $P(X)$ is false [true]. The same
definitions of validity and completeness can be used to show that
consistency is a consequence of validity and that the system is
incomplete if the sentences $PN(\sim PN),\; \sim PN(\sim PN)$ are
included.

This interpretation is much simpler than the one examined in this
paper as truth is defined everywhere and $\sim P(X)$ is true
[false] if $P(X)$ is false [true]. However it makes no use of the
quantum mechanical nature of the enumeration system. Also the
path sum description of the evolution plays no role in this
interpretation as there is no path connection between the
occurrence of $P(X)$ or $\sim P(X)$ and $X$.

The existence of different interpretations or models of the
sentences is well known in mathematical logic in that consistent
axiom systems have many different inequivalent models
\cite{Shoenfield,Frankel}.  For some axiom systems some models
are more useful than others.  An example is arithmetic where the
standard model is almost universally used.  However there also
exist many nonstandard models of arithmetic which may be useful
for some purposes.

In quantum mechanics the freedom of choice of interpretations is
much greater that in classical mechanics. For example there are
many linear combinations of the five basis states $|P\rangle ,\;
|(\rangle ,\; |)\rangle ,\; |\sim \rangle ,\; |0\rangle$ that
also can be used to represent the five symbols. In addition the
choice of linear combinations can be different at different
lattice sites.

This freedom is similar to the gauge freedom that exists in
quantum field theory in that many different gauge choices are
possible \cite{QFT}.  This similarity may be quite important in
future developments of the ideas presented here.

In spite of the specialized nature of the example, it does serve
to introduce the use of mathematical logical concepts such as
truth, validity, consistency, and completeness into physics in a
fashion similar to how they are defined and used in mathematical
logic \cite{Shoenfield,Frankel}. It is strongly suspected that
the definitions of these concepts and their use as restrictions
on the generators of the dynamics of a sentence generating system
is quite general and applies to the situation where the system
that enumerates sentences is distinct from the system to which
the sentences refer. In this context the existence question is
very important.

It is also suspected that the classical limit of these systems
may play an important role.  However the classical limit must be
taken so that at each step in the limiting process the dynamics
generator refers to its own validity and completeness to the
maximum extent possible.   It is speculated that this type of limit of 
quantum robots interacting with environments may have some
important and significant characteristics of intelligence.  As
such these limit systems may be quite different from classical computers 
which are also limits of quantum systems.

\section*{Acknowledgements}
This work is supported by the U.S. Department of Energy, Nuclear 
Physics Division, under contract W-31-109-ENG-38.

\section*{Appendix}

In the main text n,m-truth and n,m-validity were defined. The value $m=0$ was
chosen and the limit $n\rightarrow \infty$ of n,0-truth was used
to define truth and validity.  Another way to generate limit
definitions is to start with n,m-truth and take the limit
$m\rightarrow \infty$ to define n-truth and n-validity.  Truth
and validity are then defined by taking the limit $n\rightarrow
\infty$.  The independence of the limit definitions on the choice
of $m$ follows if it can be proved that the two limit definitions
are the same.  That is one must prove that
\begin{equation}
\lim_{n\rightarrow \infty}\lim_{m\rightarrow \infty}\langle \Psi
(n)|(T^{\dagger})^{m}Q^{h}_{X_{S}}T^{m}Q^{h}_{S}| \Psi
(n)\rangle =\lim_{n\rightarrow \infty}\langle \Psi
(n)|Q^{h}_{X_{S}}Q^{h}_{S}| \Psi (n)\rangle \label{equal}
\end{equation}

To see that this is the case, note that the following relations
hold: 
\begin{displaymath}
\langle \Psi (n)|(T^{\dagger})^{m}Q^{h}_{X_{S}}
T^{m}Q^{h}_{S}| \Psi (n)\rangle =\langle \Psi
(n)|(T^{\dagger})^{m}Q^{h}_{X_{S}} Q^{h}_{S,[0,n-2],m} T^{m}|
\Psi (n)\rangle \leq \langle
\Psi(n+m)|Q^{h}_{X_{S}}Q^{h}_{S}|\Psi (n+m)\rangle.
\end{displaymath}
where Eqs. \ref{Qcomm} and \ref{Qltgt} have been used along with
the commutativity of $Q^{h}_{X_{S}}$ and $Q^{h}_{S}$. Since this
is true for each $n,m$ the lefthand limit is $\leq$ the righthand
limit.

Conversely for each $n$ the unitarity of $T$ and the above noted
commutativity and referenced equations give
\begin{displaymath}
\langle \Psi(n)|Q^{h}_{X_{S}}Q^{h}_{S}|\Psi (n)\rangle = \langle
\Psi(n)|(T^{\dagger})^{m}Q^{h}_{X_{S,[0,n-
2],m}}T^{m}Q^{h}_{S}|\Psi (n)\rangle \leq  \langle
\Psi(n)|Q^{h}_{S}(T^{\dagger})^{m}Q^{h}_{X_{S}}T^{m}
Q^{h}_{S}|\Psi (n)\rangle
\end{displaymath}
which completes the proof. It follows from this that Eq. \ref{equal} also 
holds if $Q^{h}_{X_{S}}$ is replaced by $Q^{h}_{\neg X_{S}}$.

\end{document}